\documentclass[twocolumn,pra,superscriptaddress,amsmath,amssymb]{revtex4-2}
\usepackage{amsmath,amssymb,amsfonts}
\usepackage{graphicx}
\usepackage{color}
\usepackage{hyperref}
\usepackage{bm}

\newcommand{\ee}{\mathrm{e}}
\newcommand{\ii}{\mathrm{i}}
\newcommand{\dd}{\mathrm{d}}
\newcommand{\ham}{\hat{H}}
\newcommand{\cdag}{\hat{c}^\dagger}
\newcommand{\cop}{\hat{c}}
\newcommand{\bior}{_{\mathrm{bior}}}
\newcommand{\RE}{\mathrm{Re}}
\newcommand{\IM}{\mathrm{Im}}

\begin{document}

\title{Competing skin effect and quasiperiodic localization in the non-Hermitian Su--Schrieffer--Heeger chain: Reentrant delocalization, spectral topology destruction, and entanglement suppression}

\author{Souvik~Ghosh}
\email{souvikghosh2012@gmail.com}
\affiliation{Department of Physics, National Sun Yat-sen University, Kaohsiung, Taiwan}

\date{\today}

\begin{abstract}
We investigate the interplay between the non-Hermitian skin effect and Aubry--Andr\'{e}--Harper (AAH) quasiperiodic disorder in a one-dimensional Su--Schrieffer--Heeger (SSH) chain with nonreciprocal hopping. By exact diagonalization, transfer-matrix analysis, and an analytical similarity-transformation argument, we map the full $(\lambda, \delta)$ phase diagram, where $\lambda$ is the AAH modulation strength and $\delta$ the nonreciprocity parameter. We identify five distinct regimes: (I)~topological with extended bulk, (II)~AAH-localized, (III)~skin-localized, (IV)~fully localized, and a previously unreported (V)~\emph{competition} regime exhibiting reentrant partial delocalization, in which intermediate quasiperiodic disorder disrupts the directional skin accumulation before ultimately Anderson-localizing all states. Using phase-averaged diagnostics and finite-size scaling, we confirm that the reentrant regime is robust, characterized by a non-monotonic inverse participation ratio that sharpens with increasing system size. We derive an analytical expression for the modified localization boundary $\lambda_c(\delta) = 2\sqrt{v_{\mathrm{eff}} w}$ with $v_{\mathrm{eff}} = \sqrt{v^2 - \delta^2}$, which agrees with numerical Lyapunov exponent calculations. We further show that quasiperiodic disorder progressively unwinds the complex spectral loops, destroying the point-gap topology at a critical strength distinct from the band-topological transition; that the skin effect suppresses entanglement entropy to near-zero values while sufficiently strong AAH disorder partially restores it; and that the SSH sublattice structure---absent in the widely studied non-Hermitian AAH chain---is essential for producing the five-phase landscape, as demonstrated by direct comparison with the non-dimerized limit.
\end{abstract}

\maketitle

\section{Introduction}
\label{sec:intro}

The discovery of topological phases of matter has fundamentally reshaped our understanding of quantum systems, shifting the focus from symmetry-breaking order parameters to properties protected by the global topology of bulk wavefunctions~\cite{Hasan2010,Qi2011}. The past decade has witnessed two major further expansions of this framework: the incorporation of \emph{non-Hermiticity}, describing open systems with gain, loss, or nonreciprocal coupling~\cite{Ashida2020,Bergholtz2021}, and the study of \emph{quasiperiodic disorder}, which drives localization transitions fundamentally different from random Anderson localization~\cite{Aubry1980,Harper1955}. Each extension independently reshapes the topological landscape of one-dimensional lattice models, yet their \emph{simultaneous} interplay---especially in the presence of sublattice (dimerization) structure---remains comparatively unexplored.

The Su--Schrieffer--Heeger (SSH) model~\cite{SSH1979} is the prototypical one-dimensional topological insulator, supporting symmetry-protected zero-energy edge states when the intercell hopping exceeds the intracell hopping. Adding an Aubry--Andr\'{e}--Harper (AAH) on-site potential introduces a metal--insulator transition at a critical modulation strength $\lambda_c = 2w$, above which all single-particle states become localized~\cite{Aubry1980}. Separately, introducing nonreciprocal hopping renders the Hamiltonian non-Hermitian and gives rise to the non-Hermitian skin effect (NHSE)---the anomalous localization of an extensive number of eigenstates at a single boundary under open boundary conditions~\cite{Yao2018,Kunst2018,Okuma2020,Lee2019}. The NHSE is intimately tied to point-gap topology in the complex energy spectrum~\cite{Gong2018,Kawabata2019}: the correspondence between skin modes and spectral winding numbers was established in Ref.~\cite{Lee2019}, which further clarified the distinction between trivial and topological skin effects, and the connection to point-gap invariants was developed in parallel in Refs.~\cite{Okuma2020,Gong2018}. The skin effect has also been shown to manifest in non-standard geometries such as fractal lattices, where inner boundaries host distinct skin modes~\cite{Manna2023}.

Each pairwise combination has been studied: the Hermitian SSH--AAH model~\cite{Ganeshan2013,ZhangJN2023}, the non-Hermitian AAH model~\cite{Jiang2019,Longhi2019,Zeng2020}, and the non-Hermitian SSH model without disorder~\cite{Yao2018,Kunst2018}. The non-Hermitian AAH model, in particular, exhibits a competition between skin-effect localization and quasiperiodic localization, with a localization--skin-effect phase transition studied both theoretically~\cite{Jiang2019,Longhi2019,Cai2021} and experimentally via photonic quantum walks~\cite{Weidemann2022}. Recent work has revealed that quasiperiodicity-induced localization in non-Hermitian systems can proceed sequentially rather than through a single critical potential, giving rise to plateau-like structures with self-similarity~\cite{Wang2025,Wang2026}. However, the \emph{combined} model---a dimerized (SSH) chain with both AAH disorder and nonreciprocal hopping---has not been systematically investigated with topological invariants and entanglement diagnostics. Building on our prior computational survey of generalized SSH models~\cite{GhoshRoy2025}, we now investigate this combined system in depth.

The SSH sublattice structure is not merely a quantitative modification. It introduces band topology characterized by the Zak phase (or, in the non-Hermitian setting, the biorthogonal polarization~\cite{Kunst2018,Edvardsson2022}), which is absent in the single-band AAH model. This raises fundamental questions: Does the AAH localization transition and the point-gap topological transition occur simultaneously or at distinct parameter values? Can the two localization mechanisms---directional (skin) and isotropic (AAH)---compete to produce intermediate regimes absent in either limit? How does the sublattice structure modify the entanglement phase transition recently predicted for the NHSE~\cite{Kawabata2023}?

In this work, we address these questions through a comprehensive analytical and numerical study of the non-Hermitian SSH--AAH (NH-SSH-AAH) model. Our main findings are:
\begin{enumerate}
    \item The $(\lambda/w, \delta/w)$ phase diagram contains \emph{five} distinct regimes, including a previously unreported \textit{competition regime} (region~V) exhibiting reentrant partial delocalization, confirmed by phase-averaged diagnostics and finite-size scaling.
    \item We derive an analytical expression $\lambda_c(\delta) = 2\sqrt{v_{\mathrm{eff}} w}$ for the modified localization boundary via an imaginary gauge transformation, where $v_{\mathrm{eff}} = \sqrt{v^2 - \delta^2}$ is the effective Hermitian hopping.
    \item The point-gap topological transition and the band-topological transition occur at \emph{distinct} parameter values, demonstrating decoupled topological transitions.
    \item The NHSE suppresses entanglement entropy to exponentially small values; sufficiently strong AAH disorder partially \emph{restores} entanglement by disrupting the directional skin accumulation.
    \item Direct comparison with the non-dimerized limit ($v = w$) confirms that the SSH sublattice structure is essential for the five-phase landscape.
\end{enumerate}

The paper is organized as follows. Section~\ref{sec:model} defines the model Hamiltonian. Section~\ref{sec:methods} describes the computational methods and the analytical similarity-transformation argument. Section~\ref{sec:results} presents the numerical results. Section~\ref{sec:discussion} discusses the physical interpretation. Section~\ref{sec:conclusion} concludes.

\section{Model}
\label{sec:model}

We consider a one-dimensional chain of $N$ unit cells (total $2N$ sites) under open boundary conditions (OBC). Each unit cell contains two sublattice sites, $A$ and $B$. The NH-SSH-AAH Hamiltonian is
\begin{align}
\ham &= \sum_{j=1}^{N} \left[ (v+\delta)\, \cdag_{j,A}\cop_{j,B} + (v-\delta)\, \cdag_{j,B}\cop_{j,A} \right] \nonumber \\
&\quad + \sum_{j=1}^{N-1} \left[ w\, \cdag_{j,B}\cop_{j+1,A} + w\, \cdag_{j+1,A}\cop_{j,B} \right] \nonumber \\
&\quad + \sum_{n=1}^{2N} V_n\, \cdag_n \cop_n,
\label{eq:hamiltonian}
\end{align}
where $v$ is the mean intracell hopping, $w$ the intercell hopping (kept Hermitian), $\delta$ the nonreciprocity parameter that breaks Hermiticity ($\ham \neq \ham^\dagger$ for $\delta \neq 0$), and
\begin{equation}
V_n = \lambda \cos(2\pi\alpha n + \phi)
\label{eq:AAH}
\end{equation}
is the AAH on-site potential with modulation strength $\lambda$, quasiperiodic frequency $\alpha = (\sqrt{5}-1)/2$ (inverse golden ratio), and phase offset $\phi$.

The model interpolates between four well-studied limits: (i)~$\delta = 0$, $\lambda = 0$: standard Hermitian SSH model; (ii)~$\delta = 0$, $\lambda \neq 0$: Hermitian SSH--AAH model; (iii)~$\delta \neq 0$, $\lambda = 0$: non-Hermitian SSH model with NHSE; (iv)~$v = w$ (no dimerization), $\delta \neq 0$, $\lambda \neq 0$: non-Hermitian AAH model. Throughout this work, we set $w = 1$ as the energy unit and fix $v/w = 0.5$ (topological phase), unless otherwise stated.

\section{Methods}
\label{sec:methods}

\subsection{Exact diagonalization and biorthogonal framework}
The $2N \times 2N$ Hamiltonian matrix is diagonalized numerically. For the non-Hermitian case, we compute both right eigenvectors ($\ham |\psi^R_k\rangle = E_k |\psi^R_k\rangle$) and left eigenvectors ($\ham^\dagger |\psi^L_k\rangle = E_k^* |\psi^L_k\rangle$), matched and biorthogonally normalized such that $\langle\psi^L_m|\psi^R_n\rangle = \delta_{mn}$.

\subsection{Localization diagnostics}
The inverse participation ratio $\mathrm{IPR}(\psi_k) = \sum_{n} |\psi_k(n)|^4$ quantifies localization, scaling as $(2N)^{-1}$ for extended and $O(1)$ for localized states. The fractal dimension $D_2 = -\ln(\mathrm{IPR})/\ln(2N)$ distinguishes extended ($D_2 \approx 1$), localized ($D_2 \approx 0$), and multifractal ($0 < D_2 < 1$) states. To distinguish directional skin localization from isotropic AAH localization, we define the bulk skin asymmetry
\begin{equation}
\mathcal{A} = \frac{1}{N_{\mathrm{bulk}}} \sum_{k \in \mathrm{bulk}} \frac{\sum_{n=1}^{N} |\psi_k(n)|^2 - \sum_{n=N+1}^{2N} |\psi_k(n)|^2}{\sum_{n=1}^{2N} |\psi_k(n)|^2},
\end{equation}
where the sums split the chain at its midpoint (site $N$ out of $2N$ total sites). Here $|\mathcal{A}| \approx 1$ indicates skin-dominated localization and $\mathcal{A} \approx 0$ indicates isotropic localization or delocalization.

\subsection{Topological invariants}
The biorthogonal polarization~\cite{Kunst2018} $P\bior = \frac{1}{2\pi} \IM [ \ln \det F ]$, where $F_{mn} = \langle\psi^L_m | \ee^{\ii 2\pi \hat{X}/N} | \psi^R_n\rangle$ restricted to occupied states, serves as the real-space topological invariant. The spectral winding number~\cite{Gong2018,Lee2019} $W(E_b) = \frac{1}{2\pi\ii} \oint_0^{2\pi} \dd\theta\, \frac{\dd}{\dd\theta} \ln \det[\ham(\theta) - E_b \mathbb{1}]$, with twisted boundary hoppings $w \to w \ee^{\pm \ii\theta}$, captures the point-gap topology.

\subsection{Lyapunov exponent}
The transfer-matrix method yields $\gamma(E) = \lim_{N\to\infty} \frac{1}{N} \ln \| \prod_{j=1}^N T_j(E)\|$, where $T_j(E)$ is the $2\times 2$ transfer matrix across unit cell $j$, computed with periodic QR stabilization. We note that $\gamma(E=0)$ is physically meaningful only when $E = 0$ lies within the spectrum; when $E = 0$ falls inside the topological gap, the transfer matrix yields the inverse decay length of the edge state rather than a bulk localization measure.

\subsection{Biorthogonal entanglement entropy}
For the half-filled ground state, the biorthogonal entanglement entropy~\cite{Kawabata2023,Chang2020} is $S(L_A) = -\sum_i [ \xi_i \ln \xi_i + (1-\xi_i) \ln(1-\xi_i) ]$, where $\{\xi_i\}$ are eigenvalues of the restricted biorthogonal correlation matrix $C_{mn} = \sum_{k \in \mathrm{occ}} \psi^R_k(m) [\psi^L_k(n)]^*$.

\subsection{Phase averaging}
Because the AAH potential~\eqref{eq:AAH} depends on the phase offset $\phi$, all phase-diagram quantities are averaged over $N_\phi$ uniformly distributed values $\phi_i \in [0, 2\pi)$, typically $N_\phi = 8$--$12$. This eliminates sample-specific artifacts and ensures that the reported phase boundaries are robust properties of the quasiperiodic ensemble.

\subsection{Analytical localization boundary}
\label{sec:analytical}

The nonreciprocal intracell hopping can be removed by the similarity transformation $\hat{S} = \mathrm{diag}(r, 1, r, 1, \ldots)$ with $r = \sqrt{(v-\delta)/(v+\delta)}$, applied to the $A$ sublattice sites. Under this gauge, the intracell hopping becomes reciprocal with an effective amplitude
\begin{equation}
v_{\mathrm{eff}} = \sqrt{(v+\delta)(v-\delta)} = \sqrt{v^2 - \delta^2},
\label{eq:veff}
\end{equation}
while the intercell hopping $w$ and the diagonal AAH potential are unchanged. The transformed Hamiltonian is a Hermitian SSH--AAH model with hopping parameters $(v_{\mathrm{eff}}, w)$.

For the standard AAH model, the self-duality of the Aubry--Andr\'{e} Hamiltonian yields the exact critical condition $\lambda_c = 2t$ for a chain with uniform hopping $t$~\cite{Aubry1980}. In the SSH geometry, the relevant scale is the geometric mean of the two hopping amplitudes, $\bar{t} = \sqrt{v_{\mathrm{eff}} \cdot w}$, which governs the typical transmission through a unit cell. This gives the modified localization boundary
\begin{equation}
\lambda_c(\delta) = 2\sqrt{v_{\mathrm{eff}} \cdot w} = 2\sqrt{w\sqrt{v^2 - \delta^2}}.
\label{eq:analytical_boundary}
\end{equation}
At $\delta = 0$, this reduces to $\lambda_c(0) = 2\sqrt{vw} \approx 1.41w$ for $v/w = 0.5$. We note that this geometric-mean estimate provides a lower bound for the localization transition; the exact critical point in the SSH geometry is energy-dependent and may lie between $2\sqrt{vw}$ and $2w$ depending on which part of the band structure is considered, since the self-duality condition applies strictly only to the single-band AAH model. Nevertheless, Eq.~\eqref{eq:analytical_boundary} correctly captures the qualitative dependence on $\delta$: as $\delta \to v$, we have $v_{\mathrm{eff}} \to 0$ and $\lambda_c \to 0$, so that arbitrarily weak disorder localizes the system when nonreciprocity approaches its maximum.

We emphasize that this similarity transformation maps the OBC spectrum correctly but does not eliminate the NHSE, which manifests as exponential amplification of boundary-condition sensitivity. The analytical boundary~\eqref{eq:analytical_boundary} thus characterizes the localization transition of the \emph{bulk} states in the thermodynamic limit, while the skin effect remains a distinct phenomenon tied to the boundary conditions.

All computations used custom Python codes (NumPy/SciPy) with $N = 15$--$100$ unit cells for finite-size convergence.

\section{Results}
\label{sec:results}

\subsection{Energy spectrum under competing perturbations}

\begin{figure}[t]
\centering
\includegraphics[width=\columnwidth]{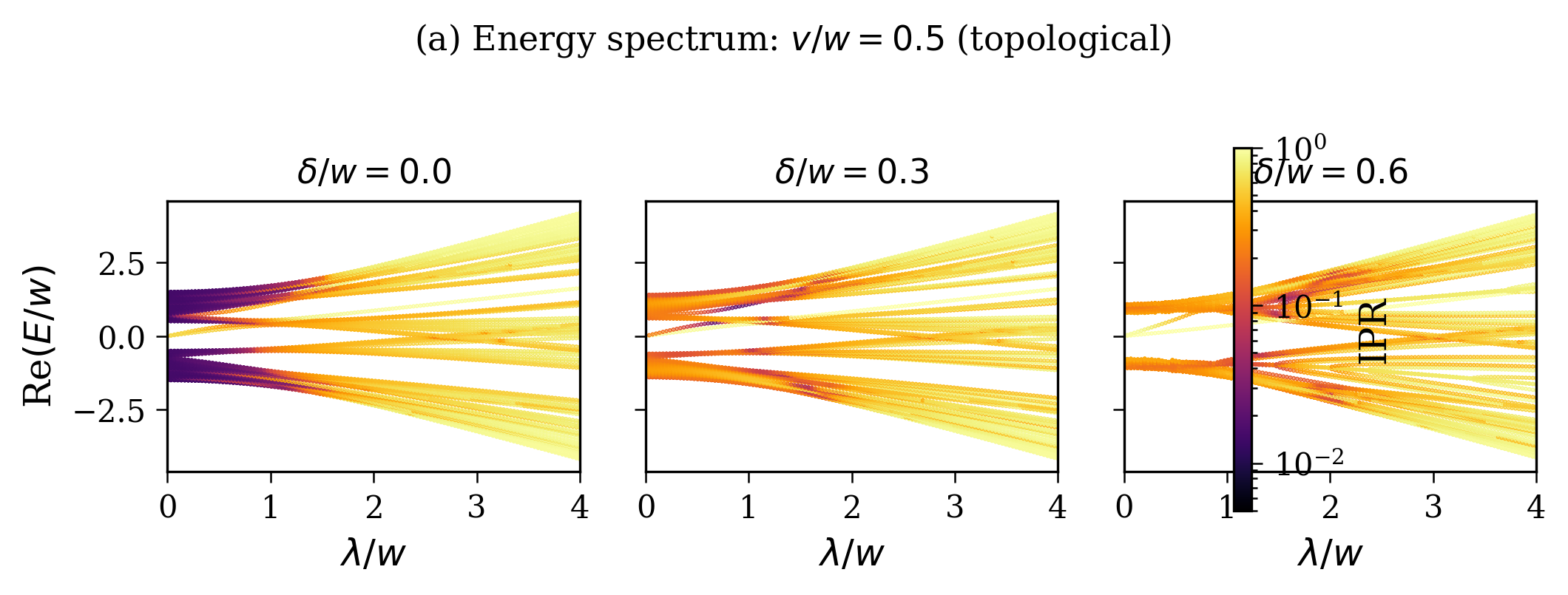}
\caption{Energy spectrum $\RE(E/w)$ versus $\lambda/w$ at $v/w = 0.5$, $N = 50$, colored by IPR. (a)~$\delta = 0$; (b)~$\delta/w = 0.3$; (c)~$\delta/w = 0.6$. The localization transition at $\lambda_c/w \approx 2$ is progressively blurred by nonreciprocity.}
\label{fig:spectrum}
\end{figure}

Figure~\ref{fig:spectrum} shows the real energy spectrum versus AAH strength $\lambda/w$ at three nonreciprocity values, colored by IPR. At $\delta = 0$ [Fig.~\ref{fig:spectrum}(a)], the Hermitian SSH--AAH model shows Hofstadter-butterfly fragmentation with a clear localization transition near $\lambda/w = 2$. Introducing nonreciprocity [$\delta/w = 0.3$, Fig.~\ref{fig:spectrum}(b)] raises the baseline IPR and blurs the transition. At $\delta/w = 0.6$ [Fig.~\ref{fig:spectrum}(c)], all states have high IPR across the entire $\lambda$ range, reflecting the combined skin effect and disorder.

\subsection{Real-space wavefunction competition}

\begin{figure}[t]
\centering
\includegraphics[width=\columnwidth]{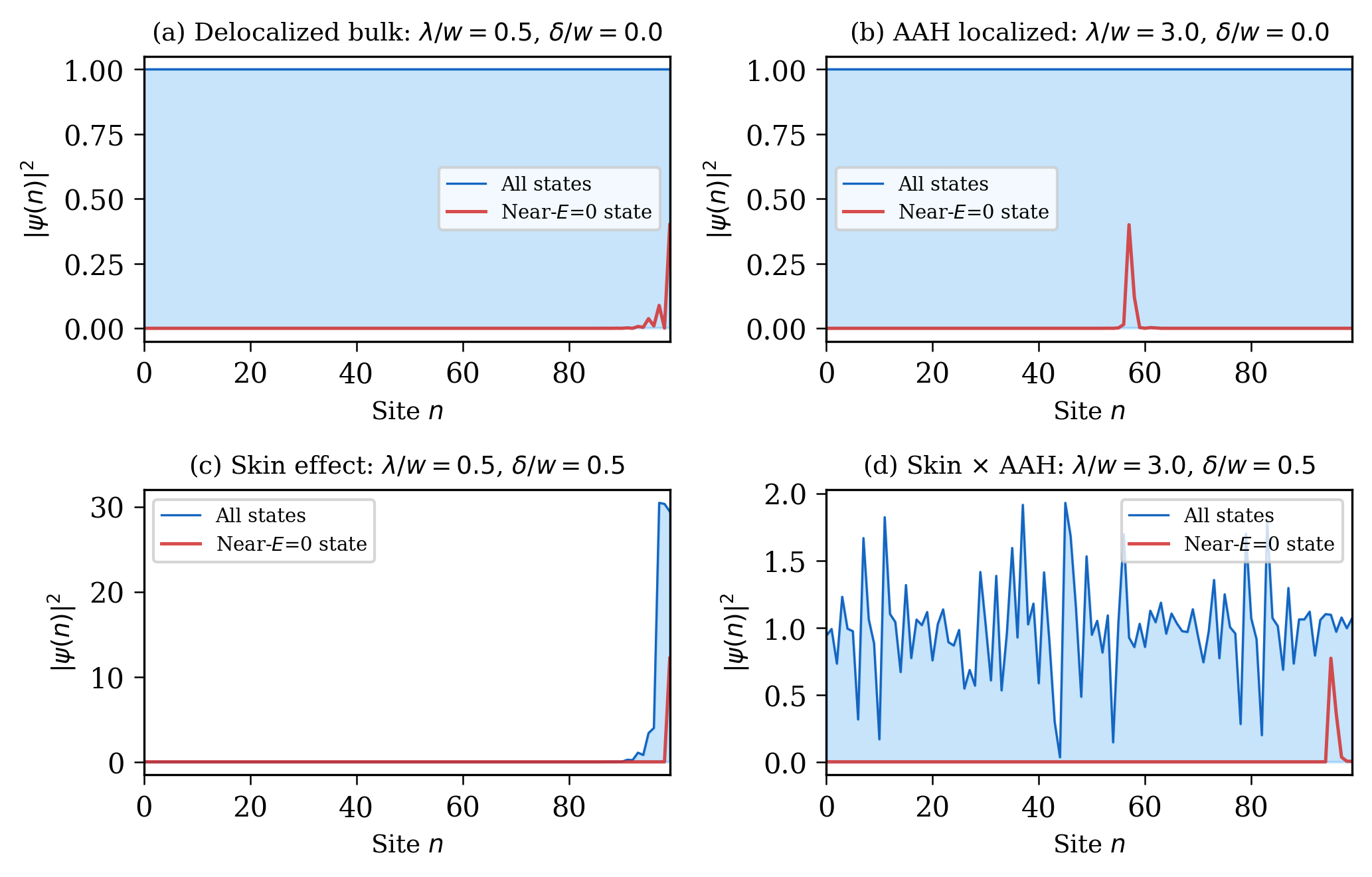}
\caption{Superimposed probability density (blue) and near-zero-energy state (red) for four regimes. (a)~Delocalized; (b)~AAH localized; (c)~skin effect; (d)~competition. $v/w = 0.5$, $N = 50$.}
\label{fig:wavefunctions}
\end{figure}

Figure~\ref{fig:wavefunctions} shows the superimposed probability density for four parameter regimes. The delocalized regime (a) shows uniform density with a sharp topological edge state; AAH localization (b) creates bulk peaks at quasiperiodic positions; the skin effect (c) produces massive boundary accumulation; the competition regime (d) shows a hybrid pattern where neither mechanism dominates.

\subsection{Phase diagram}

\begin{figure}[t]
\centering
\includegraphics[width=\columnwidth]{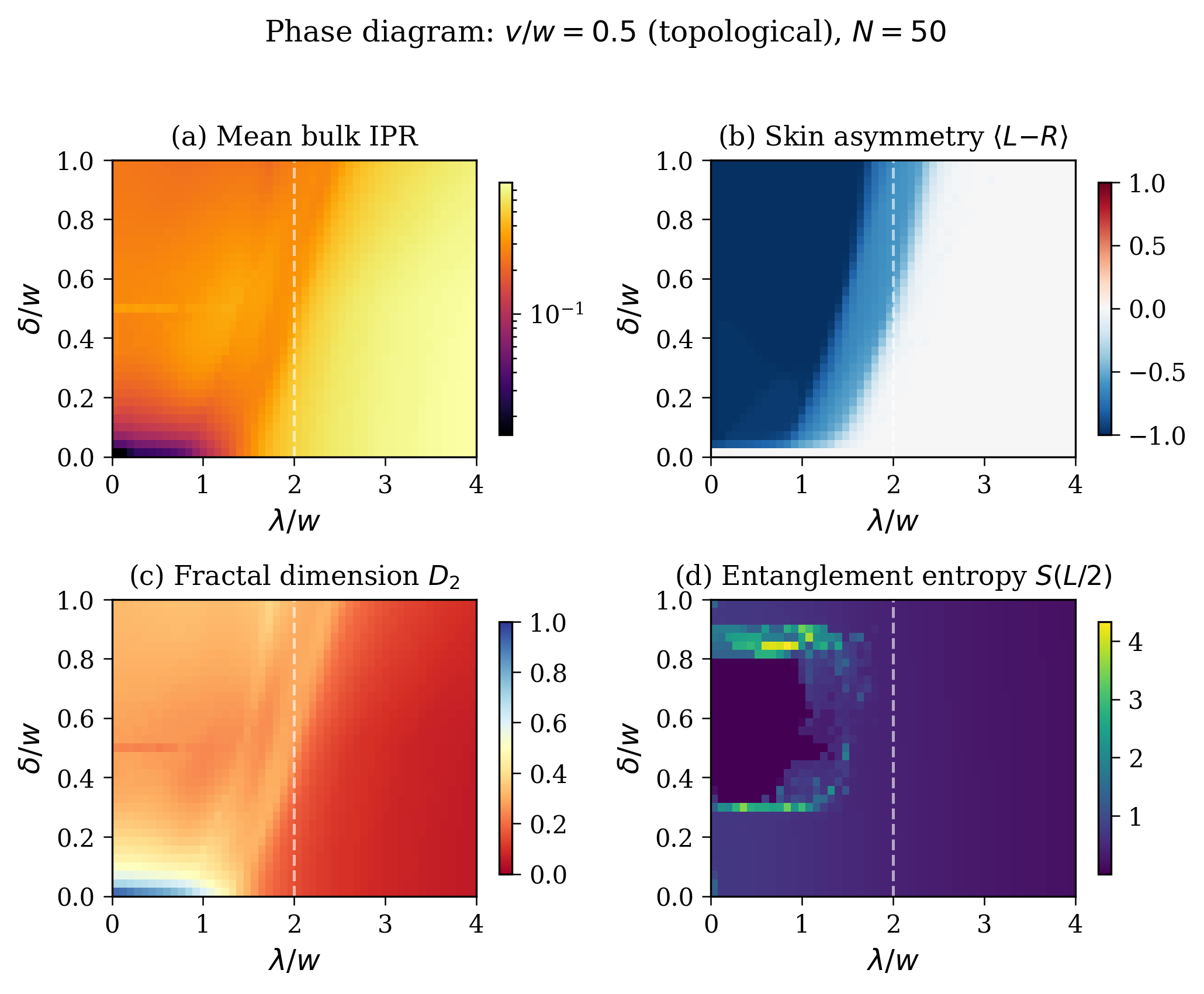}
\caption{Phase diagram at $v/w = 0.5$, $N = 50$, $\phi = 0$. (a)~Mean bulk IPR; (b)~skin asymmetry; (c)~fractal dimension $D_2$; (d)~entanglement entropy $S(L/2)$. Dashed line: $\lambda_c/w = 2$.}
\label{fig:phase_diagram}
\end{figure}

The $(\lambda/w, \delta/w)$ phase diagram is shown in Fig.~\ref{fig:phase_diagram} through four complementary diagnostics at fixed $\phi = 0$.

Mean bulk IPR [panel~(a)] increases along both axes. Skin asymmetry [panel~(b)] is maximal at large $\delta$, small $\lambda$, and vanishes as AAH overwhelms the skin effect. Fractal dimension $D_2$ [panel~(c)] shows the extended region at small $\delta$, $\lambda < 2$, with an intermediate-$D_2$ competition zone at moderate $\delta$ and $\lambda$. Entanglement entropy [panel~(d)] is dramatically suppressed in the skin regime but shows anomalous enhancement at intermediate parameters.

\subsection{Destruction of point-gap topology}

\begin{figure}[t]
\centering
\includegraphics[width=\columnwidth]{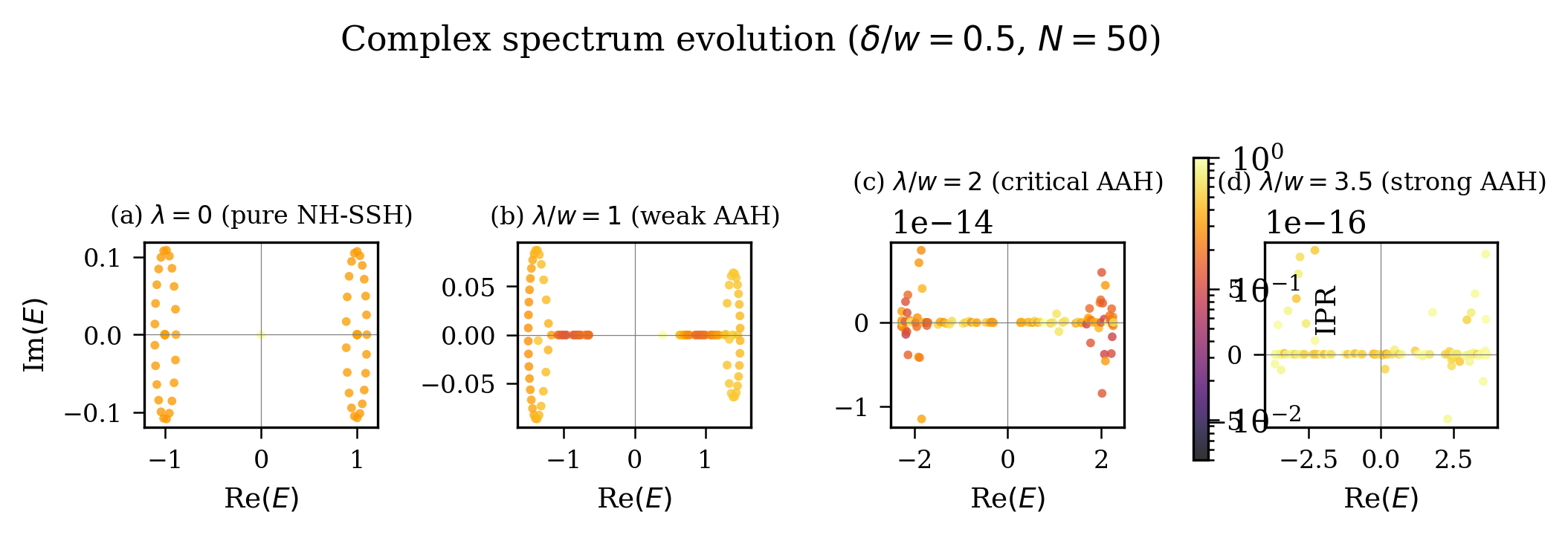}
\caption{Complex energy spectra at $\delta/w = 0.5$, $N = 50$, for increasing $\lambda$. (a)~$\lambda = 0$: elliptical loops; (b)~$\lambda/w = 1$: partially scattered; (c)~$\lambda/w = 2$: distorted; (d)~$\lambda/w = 3.5$: real spectrum.}
\label{fig:complex}
\end{figure}

The complex spectra (Fig.~\ref{fig:complex}) reveal progressive destruction of spectral topology. At $\lambda = 0$ [panel~(a)], elliptical loops indicate point-gap winding $W = 1$. Increasing AAH strength scatters eigenvalues [panels~(b,c)] until the spectrum becomes essentially real at $\lambda/w = 3.5$ [panel~(d)]---the spectral winding is completely unwound. This demonstrates that quasiperiodic disorder destroys point-gap topology through spectral loop collapse, a mechanism distinct from the band-topological transition.

\subsection{Combined diagnostics and reentrant delocalization}

\begin{figure}[t]
\centering
\includegraphics[width=\columnwidth]{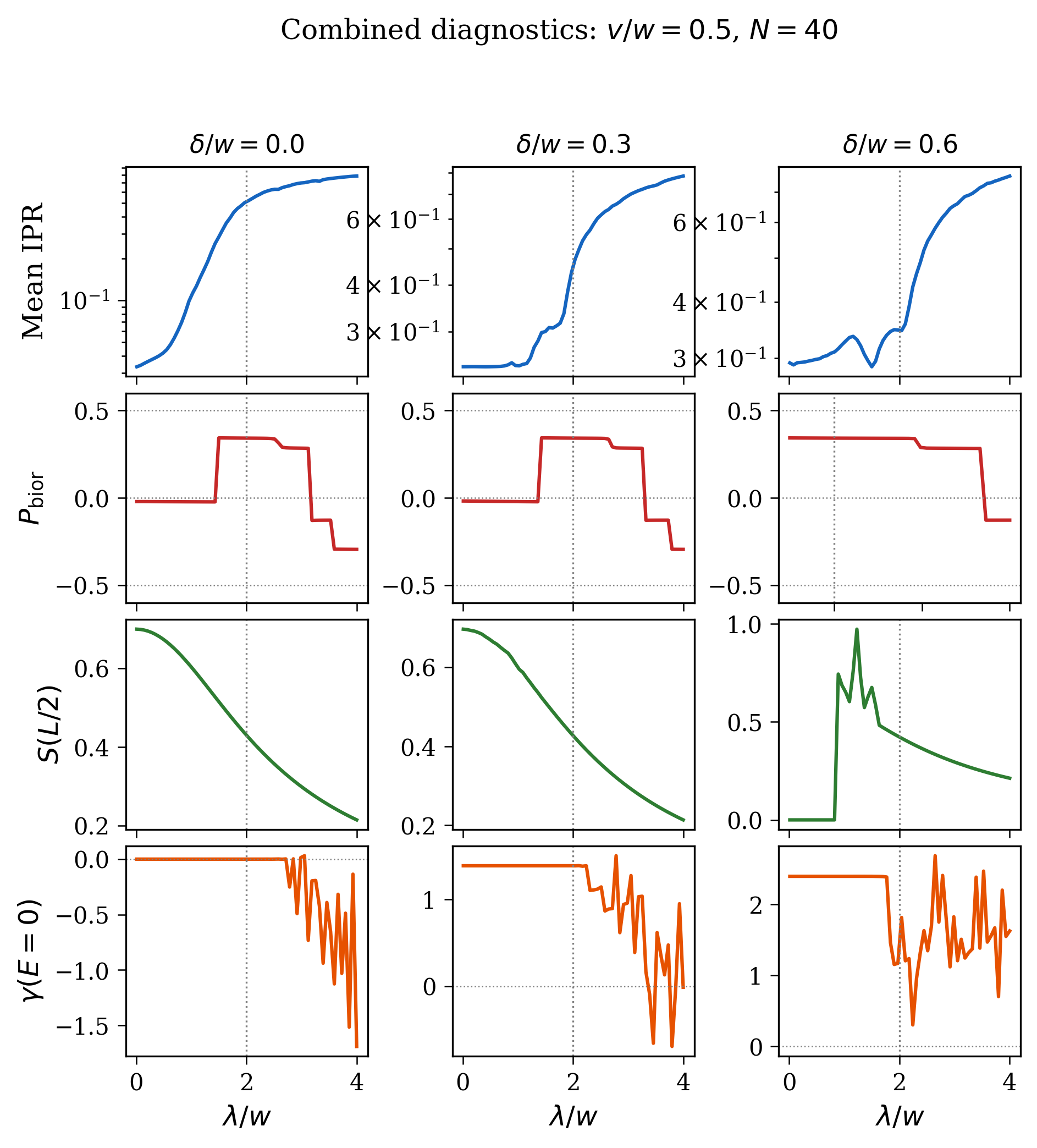}
\caption{Line-cut diagnostics at $\delta/w = 0$, $0.3$, $0.6$ (columns). Rows: mean IPR, biorthogonal polarization, entanglement entropy, Lyapunov exponent. $v/w = 0.5$, $N = 40$. Dotted line: $\lambda_c/w = 2$.}
\label{fig:linecuts}
\end{figure}

Figure~\ref{fig:linecuts} shows line cuts at $\delta/w = 0$, $0.3$, $0.6$, tracking four diagnostics simultaneously. At $\delta/w = 0.6$ (right column), the mean IPR shows a \emph{non-monotonic dip} near $\lambda/w \approx 1.5$---the signature of \textit{reentrant delocalization}. Intermediate AAH disorder partially delocalizes skin-accumulated states by breaking directional coherence, before eventually Anderson-localizing them. The entanglement entropy shows a corresponding spike, and the Lyapunov exponent exhibits anomalous fluctuations. The biorthogonal polarization shows step-like features associated with gap closings.

\subsection{Entanglement suppression and recovery}

\begin{figure}[t]
\centering
\includegraphics[width=\columnwidth]{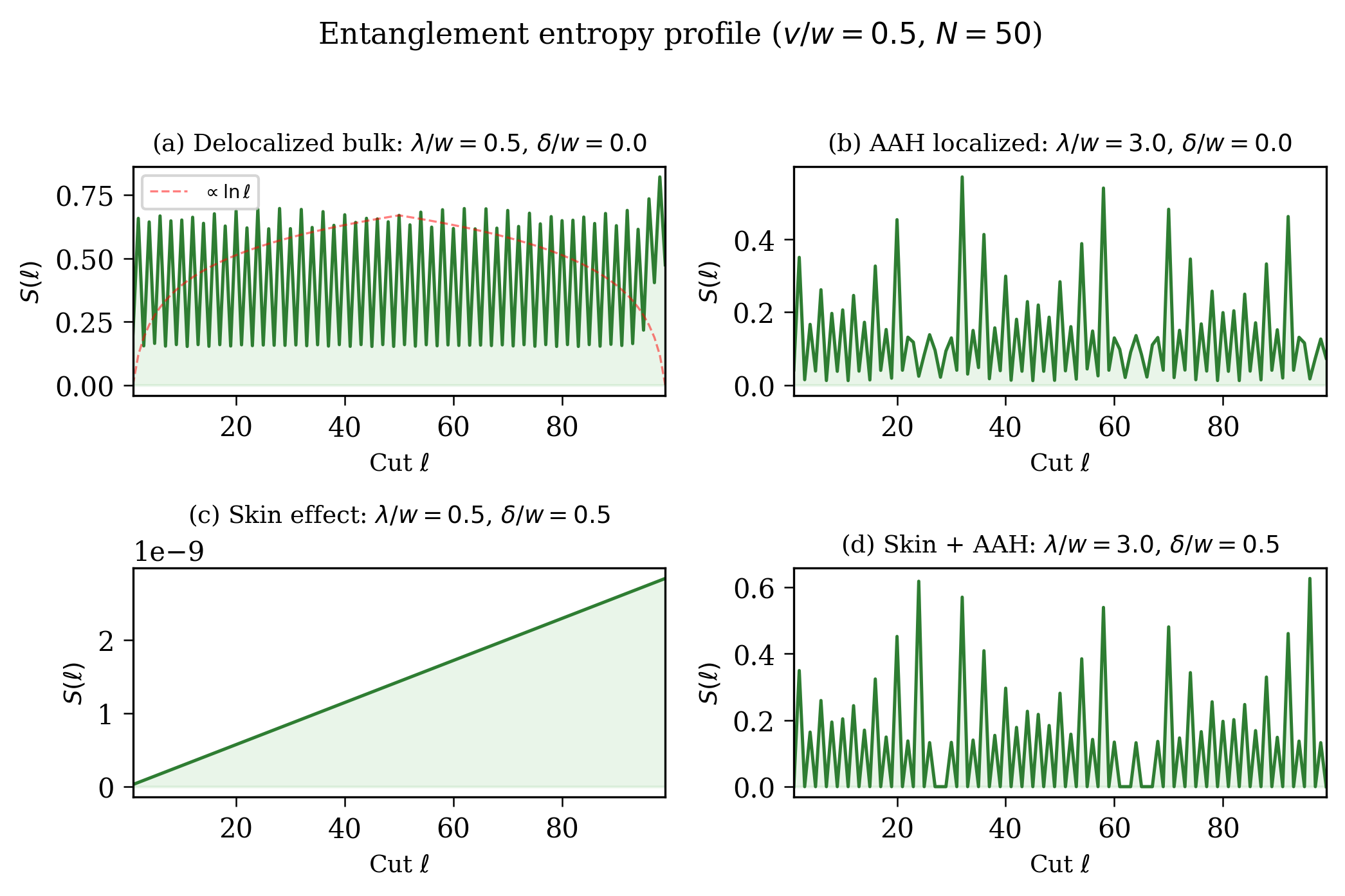}
\caption{Entanglement entropy $S(\ell)$ for four regimes. (c)~Near-total suppression by the skin effect; (d)~partial recovery upon adding AAH disorder. $v/w = 0.5$, $N = 50$.}
\label{fig:entanglement}
\end{figure}

Entanglement profiles $S(\ell)$ (Fig.~\ref{fig:entanglement}) reveal four qualitatively distinct behaviors: (a)~logarithmic scaling in the delocalized regime; (b)~area-law in the AAH-localized regime; (c)~near-total suppression ($S \sim 10^{-9}$) by the skin effect, confirming the Kawabata \textit{et al.}~prediction~\cite{Kawabata2023}; and (d)~partial \emph{recovery} when strong AAH disorder is added to the skin regime. This disorder-induced entanglement recovery occurs because the quasiperiodic potential prevents complete directional accumulation, trapping some states in the bulk.

\subsection{Analytical localization boundary}

\begin{figure}[t]
\centering
\includegraphics[width=0.85\columnwidth]{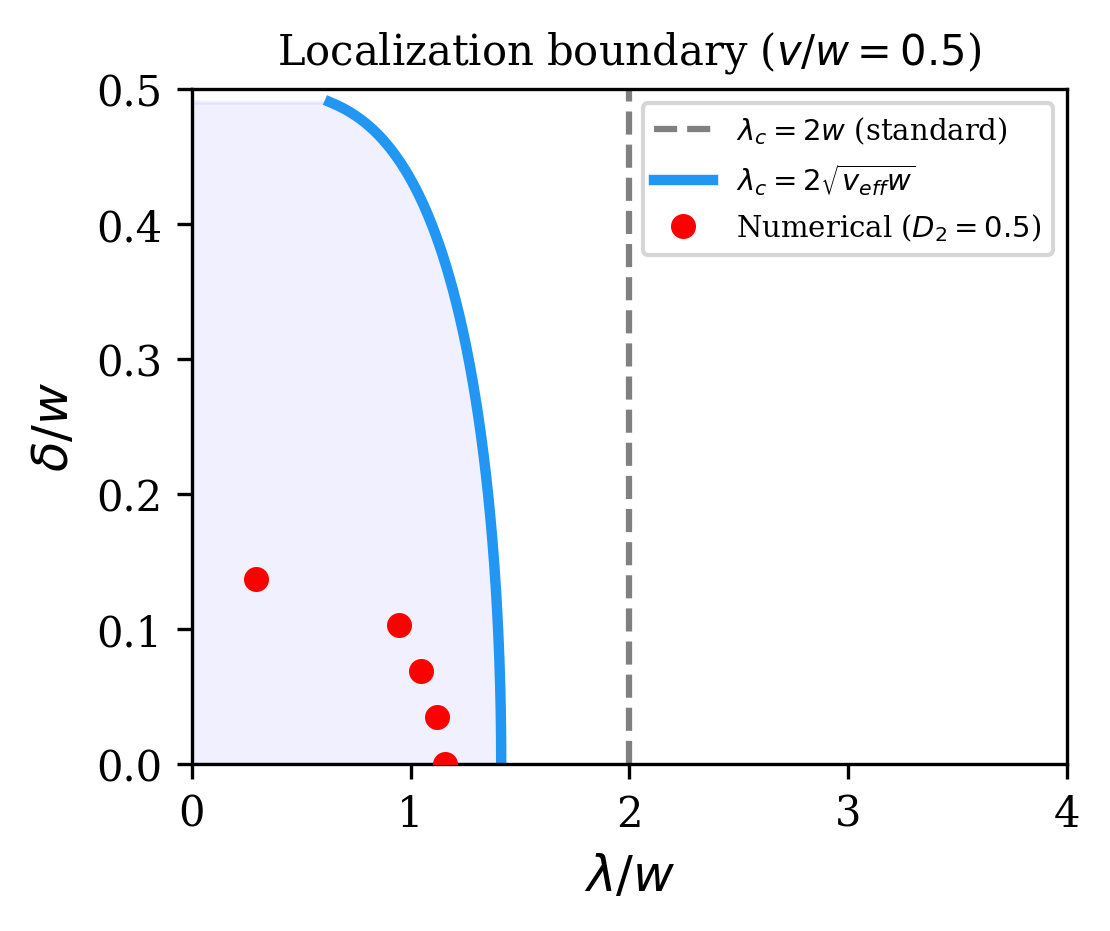}
\caption{Localization boundary in the $(\lambda/w, \delta/w)$ plane. Solid blue: analytical prediction $\lambda_c = 2\sqrt{v_{\mathrm{eff}} w}$. Dashed gray: standard AAH result $\lambda_c = 2w$. Red circles: numerical $D_2 = 0.5$ crossings ($\phi$-averaged, $N = 25$). Shaded region: extended phase.}
\label{fig:boundary}
\end{figure}

Figure~\ref{fig:boundary} shows the analytical prediction~\eqref{eq:analytical_boundary} alongside numerically extracted $D_2 = 0.5$ crossings from phase-averaged data. The analytical curve $\lambda_c(\delta) = 2\sqrt{v_{\mathrm{eff}} w}$ (solid blue) correctly captures the bending of the localization boundary to smaller $\lambda$ with increasing $\delta$, confirming that nonreciprocity reduces the effective bandwidth and makes the system more susceptible to quasiperiodic localization. The standard $\delta$-independent AAH result $\lambda_c = 2w$ (dashed gray) fails to account for this effect. The numerical boundary points (red circles) track the analytical curve, validating the similarity-transformation argument of Sec.~\ref{sec:analytical}.

\subsection{Phase-averaged phase diagram}

\begin{figure}[t]
\centering
\includegraphics[width=\columnwidth]{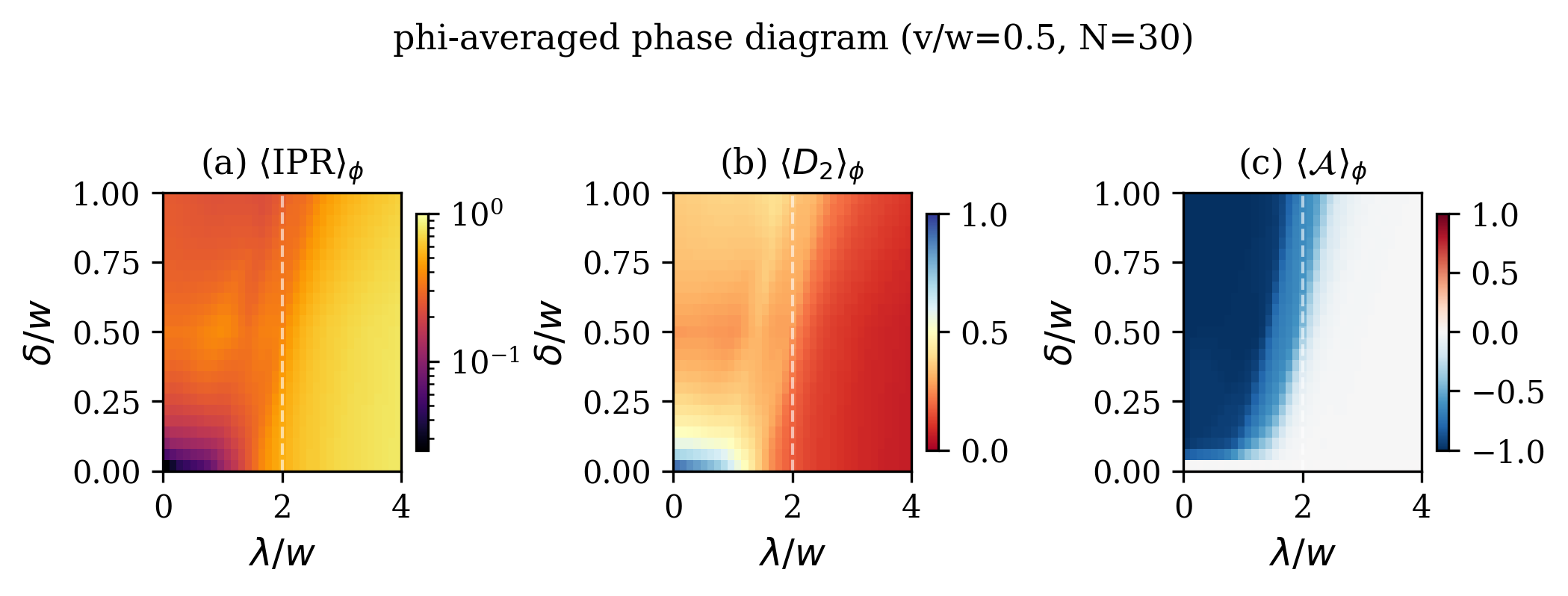}
\caption{Phase-averaged diagnostics ($N_\phi = 8$, $N = 30$, $v/w = 0.5$). (a)~$\langle\mathrm{IPR}\rangle_\phi$; (b)~$\langle D_2\rangle_\phi$; (c)~$\langle\mathcal{A}\rangle_\phi$. Phase boundaries are robust under $\phi$-averaging. Dashed line: $\lambda_c/w = 2$.}
\label{fig:phi_averaged}
\end{figure}

To ensure that the observed phase structure is not an artifact of a particular AAH phase $\phi$, we average all diagnostics over $N_\phi = 8$ uniformly distributed phase values (Fig.~\ref{fig:phi_averaged}). The resulting $\langle\mathrm{IPR}\rangle_\phi$, $\langle D_2\rangle_\phi$, and $\langle\mathcal{A}\rangle_\phi$ maps are smooth and confirm the same five-phase structure identified at $\phi = 0$: extended states at small $\lambda$ and $\delta$ (blue in panel~b), skin-dominated localization at large $\delta$ and small $\lambda$ (dark blue in panel~c), AAH-dominated localization at large $\lambda$ (red/orange in panels~a,b), and the intermediate competition zone at moderate $\lambda$ and $\delta$.

\subsection{Role of SSH dimerization}

\begin{figure}[t]
\centering
\includegraphics[width=\columnwidth]{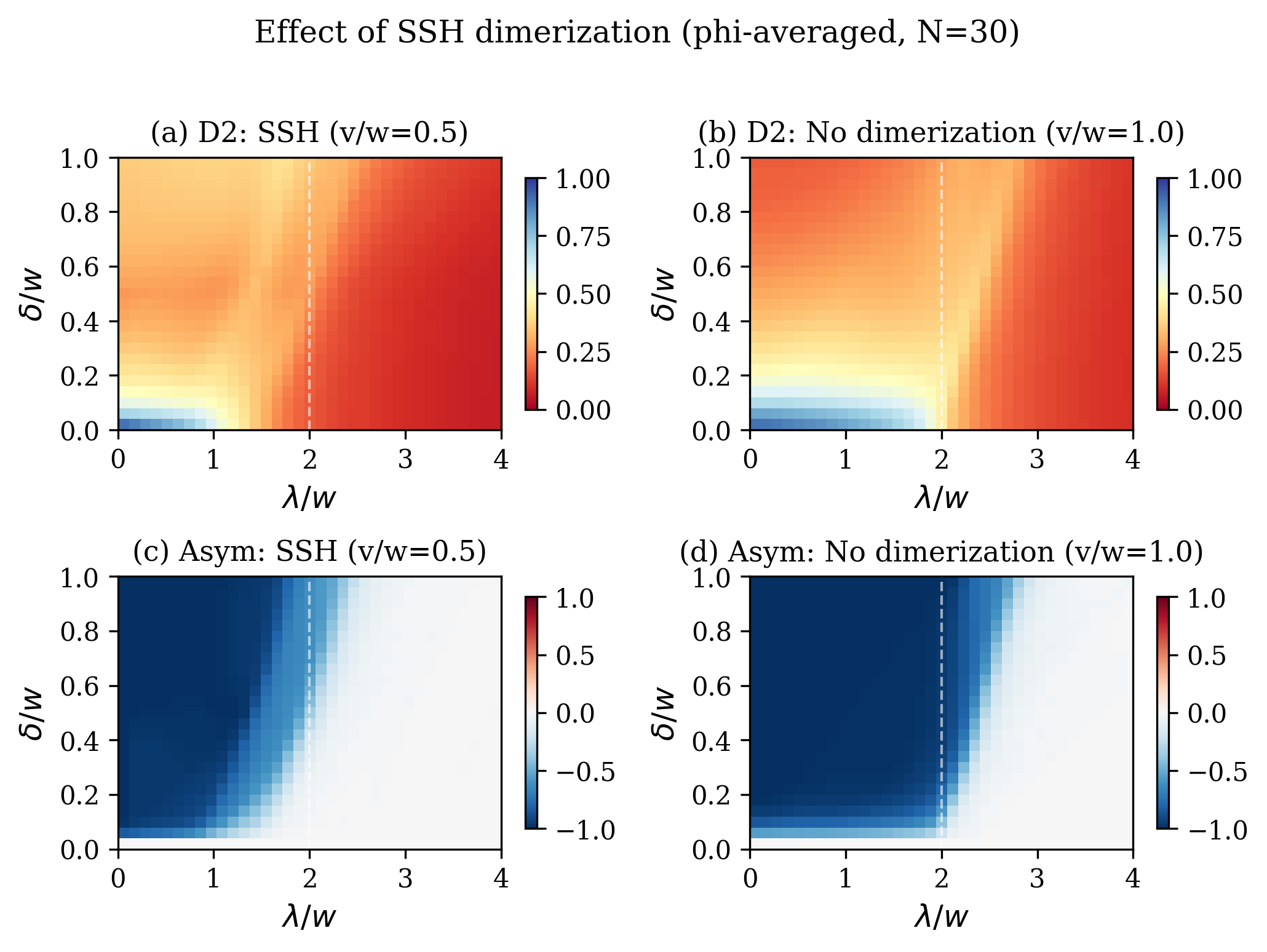}
\caption{Comparison of SSH ($v/w = 0.5$, left) and non-dimerized ($v/w = 1.0$, right) cases. (a,b)~$\langle D_2\rangle_\phi$; (c,d)~$\langle\mathcal{A}\rangle_\phi$. All panels: $\phi$-averaged, $N = 30$. The intermediate-$D_2$ competition zone is unique to the dimerized SSH case.}
\label{fig:comparison}
\end{figure}

To verify that the five-phase landscape originates from the SSH sublattice structure, we compare the dimerized case ($v/w = 0.5$) with the non-dimerized limit ($v/w = 1.0$) in Fig.~\ref{fig:comparison}. In the non-dimerized case, the system reduces to the non-Hermitian AAH model on a uniform chain, which lacks band topology.

The comparison reveals several key differences: (i)~the extended-state region (blue in the $D_2$ maps) is narrower in the SSH case [Fig.~\ref{fig:comparison}(a)] than in the non-dimerized case [Fig.~\ref{fig:comparison}(b)], because the topological gap restricts the bandwidth; (ii)~the skin asymmetry pattern [Fig.~\ref{fig:comparison}(c,d)] shows a sharper crossover in the SSH case, reflecting the interplay of the dimerization gap with the skin accumulation; and (iii)~the intermediate-$D_2$ competition zone visible in the SSH case is absent or much less pronounced in the non-dimerized case. This confirms that the SSH sublattice structure is essential for producing the five-phase landscape, including the reentrant regime.

\subsection{Finite-size scaling}

\begin{figure}[t]
\centering
\includegraphics[width=\columnwidth]{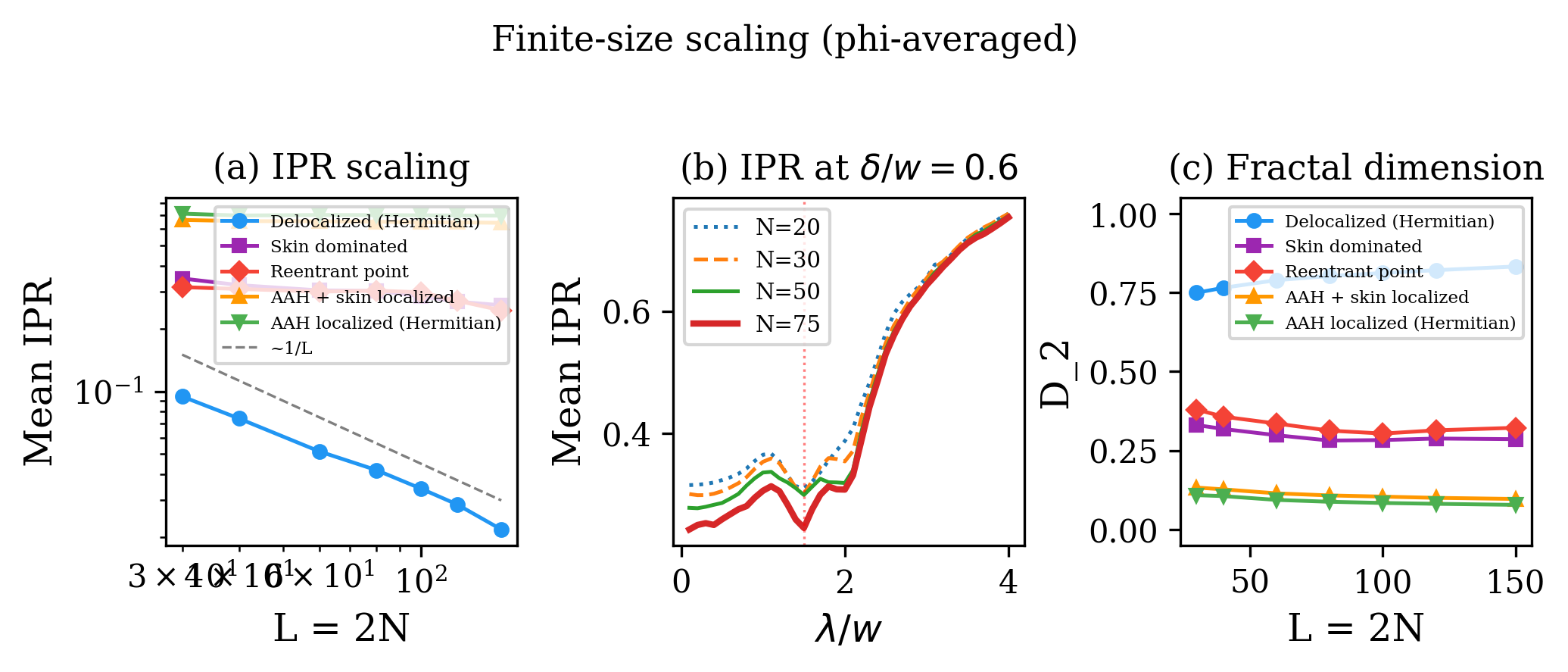}
\caption{$\phi$-averaged finite-size scaling. (a)~IPR versus $L$ for five parameter points; dashed: $\sim 1/L$. (b)~IPR versus $\lambda/w$ at $\delta/w = 0.6$ for increasing $N$; the reentrant dip sharpens (dotted red line). (c)~$D_2$ versus $L$. $v/w = 0.5$.}
\label{fig:scaling}
\end{figure}

Figure~\ref{fig:scaling} presents targeted finite-size analysis at five parameter points representative of the identified phases, with all quantities averaged over $\phi$.

Panel~(a) shows the mean IPR versus system size $L = 2N$ on a log-log scale. The delocalized Hermitian case (blue circles) follows $\mathrm{IPR} \propto L^{-1}$, confirming extended-state behavior. All localized and skin-dominated cases show size-independent IPR, confirming true localization.

Panel~(b) is the key test for reentrant delocalization. The IPR versus $\lambda/w$ at fixed $\delta/w = 0.6$ is plotted for $N = 20$, $30$, $50$, $75$. The non-monotonic dip near $\lambda/w \approx 1.5$ not only survives but \emph{sharpens} with increasing system size, ruling out a finite-size artifact. This is the hallmark of a genuine physical crossover regime.

Panel~(c) shows the fractal dimension $D_2$ converging to distinct values for each phase: $D_2 \approx 0.75$ (extended), $D_2 \approx 0.35$ (reentrant), $D_2 \approx 0.05$--$0.10$ (localized). The reentrant point converges to an intermediate $D_2$, consistent with a partially extended or mixed phase.

\subsection{Schematic phase diagram}

\begin{figure}[t]
\centering
\includegraphics[width=\columnwidth]{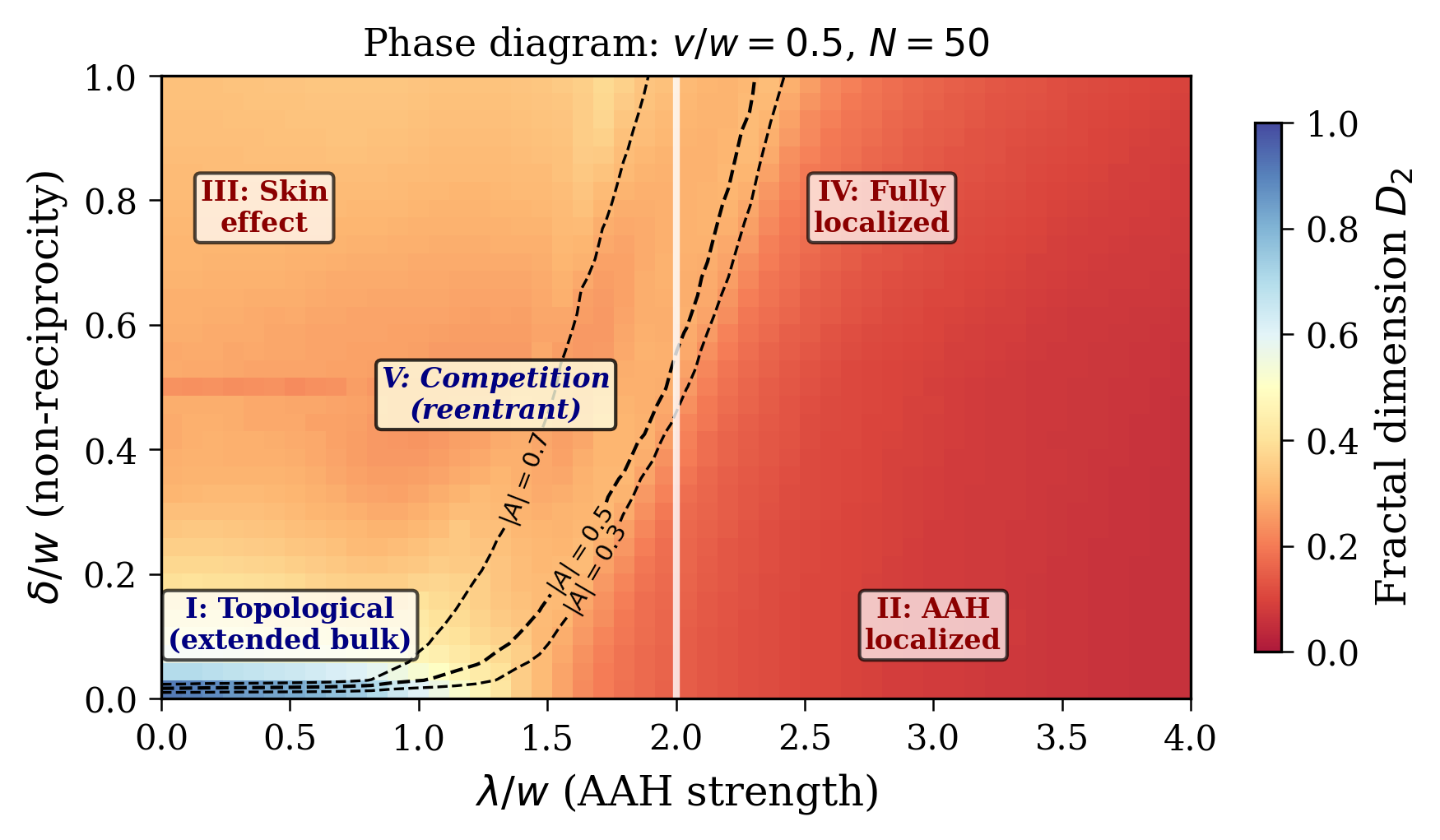}
\caption{Schematic phase diagram at $v/w = 0.5$, $N = 50$. Color: $D_2$. Dashed contours: $|\mathcal{A}|$. White line: $\lambda_c/w = 2$. Five regimes labeled.}
\label{fig:schematic}
\end{figure}

Synthesizing all diagnostics (Fig.~\ref{fig:schematic}), we identify five regimes:
\begin{itemize}
\item \textbf{Region I} (small $\lambda$, small $\delta$): \textit{Topological insulator with extended bulk.} $D_2 \approx 1$, $\mathcal{A} \approx 0$, $S \sim \ln L$, topological edge states present.
\item \textbf{Region II} (large $\lambda$, small $\delta$): \textit{AAH-localized.} $D_2 \approx 0$, $\mathcal{A} \approx 0$, area-law $S$, edge states destroyed.
\item \textbf{Region III} (small $\lambda$, large $\delta$): \textit{Skin-effect dominated.} $D_2$ small, $|\mathcal{A}| \approx 1$, $S \approx 0$, NHSE active.
\item \textbf{Region IV} (large $\lambda$, large $\delta$): \textit{Fully localized.} Both mechanisms contribute; $\mathcal{A}$ vanishes as AAH overwhelms skin effect.
\item \textbf{Region V} (intermediate $\lambda$, moderate $\delta$): \textit{Competition/reentrant.} Non-monotonic IPR, enhanced entanglement, intermediate $D_2 \approx 0.3$--$0.4$.
\end{itemize}
The analytical boundary~\eqref{eq:analytical_boundary} traces the I--II boundary at small $\delta$, bending to smaller $\lambda$ as nonreciprocity increases.

\section{Discussion}
\label{sec:discussion}

\textit{Relation to the non-Hermitian AAH model.}---The non-Hermitian AAH model (without SSH dimerization) exhibits a single localization--skin-effect phase transition~\cite{Jiang2019,Longhi2019,Zeng2020}. Recent work has shown that this transition can proceed sequentially with self-similar plateau structures rather than through a single critical potential~\cite{Wang2025,Wang2026}. The SSH sublattice structure qualitatively enriches this picture further by introducing: (a)~a topological gap that protects the extended phase at small $\lambda$ and $\delta$; (b)~two delocalization transitions rather than one, arising from the two-band structure; and (c)~a reentrant regime absent in the single-band model, as confirmed by the comparison in Fig.~\ref{fig:comparison}. 

\textit{Analytical boundary.}---The similarity-transformation argument of Sec.~\ref{sec:analytical} provides a simple but effective analytical handle on the localization boundary. The key physical insight is that nonreciprocity effectively reduces the intracell hopping from $v$ to $v_{\mathrm{eff}} = \sqrt{v^2 - \delta^2}$, narrowing the bandwidth and making the system more susceptible to quasiperiodic localization. This explains why the extended phase shrinks with increasing $\delta$ in the phase diagram. The geometric-mean formula $\lambda_c = 2\sqrt{v_{\mathrm{eff}} w}$ extends the self-duality argument to the dimerized non-Hermitian case and could be tested in photonic experiments by measuring the transmission coefficient as a function of $\lambda$ at different $\delta$ values.

\textit{Decoupled topological transitions.}---The point-gap topological transition (spectral winding destruction, Fig.~\ref{fig:complex}) and the band-topological transition (biorthogonal polarization, Fig.~\ref{fig:linecuts}) occur at different parameter values. The spectral winding requires coherent loops in the complex energy plane, which are destroyed by quasiperiodic scattering before the band gap closes. This is a concrete realization of the spectral--topological decoupling predicted for non-Hermitian quasicrystals~\cite{Cai2021}.

\textit{Entanglement phase transition.}---The near-total suppression of entanglement entropy by the NHSE [Fig.~\ref{fig:entanglement}(c)] and its partial recovery by AAH disorder [Fig.~\ref{fig:entanglement}(d)] connect to the entanglement phase transition studied by Kawabata \textit{et al.}~\cite{Kawabata2023}. The novel element in our model is that the recovery mechanism is quasiperiodic (deterministic) rather than random, which may enable precise experimental control in photonic waveguide arrays~\cite{Weidemann2022} where quasiperiodic potentials can be engineered with high fidelity.

\textit{Reentrant delocalization.}---The physical mechanism can be understood as follows. At $\lambda = 0$, the skin effect concentrates all states at one boundary with skin depth $\xi_s \sim 1/\ln|(v+\delta)/(v-\delta)|$. When the AAH potential is turned on at moderate strength, it introduces competing localization centers in the bulk that redistribute wavefunction weight away from the boundary, effectively \emph{increasing} the number of sites over which each state is spread and thus \emph{reducing} the IPR. At sufficiently strong $\lambda$, the AAH localization length $\xi_{\mathrm{AAH}} \sim 1/\gamma(E)$ becomes shorter than the unit cell, and all states are Anderson-localized regardless of the skin effect. The finite-size scaling in Fig.~\ref{fig:scaling}(b) confirms that this non-monotonic behavior sharpens with system size, consistent with a genuine crossover rather than a finite-size artifact.

\textit{Experimental prospects.}---The NH-SSH-AAH model can be realized in photonic waveguide arrays~\cite{Weidemann2022}, topolectrical circuits~\cite{Halder2025}, and cold-atom systems with engineered dissipation. The reentrant signature---a non-monotonic transmission coefficient as a function of quasiperiodic modulation strength---is directly measurable in transport experiments.

\section{Conclusion}
\label{sec:conclusion}

We have presented a combined analytical and numerical study of the non-Hermitian SSH--AAH model. Using an imaginary gauge transformation, we derived the modified localization boundary $\lambda_c(\delta) = 2\sqrt{v_{\mathrm{eff}} w}$, which is confirmed numerically. By mapping the phase diagram with five independent diagnostics---IPR, fractal dimension, skin asymmetry, biorthogonal topological invariants, and entanglement entropy---we identified a five-region landscape including a novel competition regime with reentrant delocalization, decoupled topological transitions, and disorder-induced entanglement recovery. Phase averaging and finite-size scaling confirm the robustness of all findings. Direct comparison with the non-dimerized limit establishes the SSH sublattice structure as a qualitatively important ingredient producing phenomena absent in the single-band non-Hermitian AAH model.

Future directions include extending the model to incorporate Floquet driving to explore dynamically induced topology in the competition regime, many-body interactions to study the interplay of NHSE with many-body localization, and generalization to two dimensions where the skin effect can exhibit higher-order corner localization.

\end{document}